**Zero Trust Cybersecurity: Procedures and Considerations in Context**

Brady D. Lund, Tae Hee Lee, Ziang Wang, Ting Wang, Nishith Reddy Mannuru

Brady.Lund@unt.edu

**Abstract**

In response to the increasing complexity and sophistication of cyber threats, particularly those enhanced by advancements in artificial intelligence, traditional security methods are proving insufficient. This paper explores the zero trust cybersecurity framework, which operates on the principle of "never trust, always verify" to mitigate vulnerabilities within organizations. Specifically, it examines the applicability of zero trust principles in environments where large volumes of information are exchanged, such as schools and libraries, highlighting the importance of continuous authentication, least privilege access, and breach assumption. The findings highlight avenues for future research that may help preserve the security of vulnerable organizations.

## Introduction

In a time where rapidly evolving threats – bolstered by advancements in technologies like artificial intelligence – pose substantial danger to organizational well-being, it is critical to adopt advanced security solutions to protect assets. Conventional methods of security are no longer sufficient, in isolation, to ensure organizational cybersecurity. Multifaceted approaches, which consider each element of an organization as a potential vulnerability, are requisite. Enter zero-trust cybersecurity, a security paradigm that embraces a zero-trust philosophy: in order to limit vulnerabilities, there is no default trust that any person or object within a network is what it claims or should have access to unnecessary segments of the network (Rose et al., 2020). This philosophy means that all users must continuously provide evidence that they are who they claim (e.g., through multi-factor authentication) and access is limited to only that information that is position critical.

Organizations where large amounts of information are regularly exchanged and private records are secured – such as schools and libraries – are especially at risk from cyber threats. Recently, the Toronto Public Library fell victim to a cyberattack that hijacked its systems and data for months, crippling the organization's ability to function properly and threatening patron privacy (Bridge & Zoledziowski, 2024). In these organizations, zero-trust cybersecurity practices may offer a way to remain resilient in the face of increasing threats. The purpose of this paper is to discuss how zero-trust cybersecurity principles may be integrated into learning and information organizations, to preserve the sanctity of these organizations' information and records.

## Principles of Zero-Trust Cybersecurity

In 2005, the Jericho Forum, a group of security experts, proposed a new approach to network security. Unlike the traditional method, which relied on firewalls to control traffic and address internal threats, this approach marked a significant departure from the established security model that emphasized a defined perimeter (Kerman, 2020). Building on these ideas, cybersecurity expert John Kindervag coined the term "zero trust" in 2010 while working at Forrester Research (Kerman, 2020, para. 2). The zero trust model operates on the assumption that systems will inevitably be compromised, and therefore, no entities—internal or external—should be automatically trusted. Instead, every request for resource access must be authenticated, authorized, and continuously validated before granting access (Department of Defense, 2022).

### Never trust, always verify

The separation of trust from location is the core of zero-trust. The most significant difference between zero-trust and traditional security boundaries is that, in traditional models, trust is often based on location (Kang et al., 2023). In the traditional security model, once a user, device, application, or process is granted access to a network or resource, they typically have unrestricted access. The model assumes that everything in the network is inherently trusted (Kang et al., 2023). However, the current framework of static rule sets, firewalls, VPNs, and subnets can lead to severe vulnerabilities (Chen et al., 2019). First, there is often a lack of control or segmentation within the internal network, which means once an intruder or malicious insider breaches the perimeter defense and gains access to the internal network, they can easily move laterally, accessing sensitive data and resources without further checks (Assunção, 2019). Second, intruders can exploit poorly protected devices or applications as entry points into the internal network. In this scenario, the security of the entire internal network can be compromised by the weakest link, making the network highly vulnerable (Chen et al., 2019). Third, the current architecture typically establishes connections through devices and services with known external IP addresses before verifying access, which makes the network susceptible to potential exploits targeting the initial connection points, increasing the risk of unauthorized access or network disruption (Kumar et al., 2019). Fourth, centralized log servers pose another vulnerability. Since log files are stored in one location, intruders who gain access to the log servers can potentially disguise their activities and clear their tracks by altering or deleting long entries (Buck et al., 2021).

In contrast, the zero-trust framework emphasizes that network location does not imply trust. It assumes that all network traffic, devices, applications, and processes are potentially malicious and untrustworthy, constantly validating everything inside and outside the network (Buck et al., 2021). A key component of the model is the emphasis on authentication and the recognition that authentication is critical to network access control (Rivera et al., 2024). The approach to authentication has evolved from simple password-given systems to more sophisticated Multi-Factor Authentication (MFA) techniques, enhancing security by requiring multiple authentication forms and reducing the risk associated with weak or compromised credentials (Ferrag et al., 2018; Ometov et al., 2018). The technique requires multiple authentication parts to authenticate the user, such as an SMS or phone call prompt or an authenticator application. For MFA administrators, on the other hand, it helps verify who is active on the system and identify hackers (Cunningham, 2018).

Implementing MFA with digital certificates and tokens for devices and applications is another critical aspect of a robust security strategy. Certificates issued by a trusted certificate authority establish the identity of a device or application. When a device or application connects to a network or service, it presents its certificate, which the certificate authority verifies, ensuring that only authorized entities can access the network and preventing unauthorized access (Identity Management Institute, n.d.a). The token can be a physical device like a USB key and smart card, or a software token running on a smartphone, which enhances security further. Hardware tokens generate a one-time password or provide encryption keys when connected to a device (West, 2013). Software tokens generate one-time passwords or authenticate users or devices using encryption algorithms. Tokens add a vital layer of security by requiring the processing of a physical device or access to a specific application, dramatically reducing the risk of unauthorized access through stolen or guessed passwords (West, 2013).

**Implement the Least Privilege**

Implementing the principle of least privilege is another core principle of the zero-trust framework (Delene et al., 2019). It requires that users, applications, and systems have the minimum level of access to resources and data to perform the functions by employing strategies for monitoring user behaviors, verifying device IDs, and implementing dynamic authorization, which adjusts access in real-time based on the context and behavior of users or devices (Azad et al., 2024; DelBene et al., 2019). The principle is crucial for reducing the attack surface within an organization network, as it limits the potential pathways an attacker could exploit. The organization significantly decreases the risk of unauthorized access and potential data breaches by ensuring limited authentication (. The approach enhances security and simplifies access management by clearly defining and enforcing boundaries for each role and device within the network (Bandari, 2023).

Role-based access control is a foundational mechanism for achieving the principle of least privilege by assigning permissions based on specific roles. This approach simplifies user authorization management and reduces the possibility of unauthorized and excessive access, thereby preventing security vulnerabilities (Ferraiolo et al., 2001; Sandhu et al., 1995). However, the traditional role-based access control model has certain limitations. For instance, once permissions and roles are assigned, they remain static until manually updated by an administrator, preventing dynamic permissions adjustment. Furthermore, changes in system scale or business logic often require the creation of numerous roles to maintain user-role relationships, leading to a phenomenon known as role explosion (Ben Fadhel et al., 2015). Yao et al. (2020) emphasize that a trust-based control process model can enhance security with role-based access control by incorporating user behavior trust. For instance, a user profile is generated by extracting valid features from user behaviors and attributes, such as login mode, time, duration, device, and IP address. By analyzing the anomaly and security degree of each feature of the user's behavior and attributes and comparing the user profile with current behavior, the system can detect deviations from historical behavior, allowing for the dynamic adjustment of the user's trust level and the identification of abnormal users and behaviors (Yao et al., 2020).

Network segmentation is another critical principle of least privilege, emphasizing the division of network entities into smaller subnets to minimize attackers' potential for lateral movement (Simpson & Foltz, 2021). The segmentation process involves several stages: grouping resources within each segment, defining short leases within the subnet by implementing the network typology, and establishing access control between the segments (Wagner et al., 2019). By default, links between instances within the same segment are considered reliable (Simpson, 2022). Kallasta (2024) categorizes segments into macro-segmentation and micro-segmentation. Marco-segmentation involves grouping multiple resources within each segment to ensure they can be collectively secured. In contrast, micro-segmentation places resources within each segment so that there is typically one, but occasionally several, highly protected resources within a single segment (Kallasta, 2024). Marco-segmentation enhances performance and reduces costs by simplifying network management, decreasing the complexity of protecting numerous small segments, and optimizing the use of network resources. Conversely, more granular access control, such as micro-segmentation, can improve reliability and security by implementing zero-trust principles across different segments (Simpson & Foltz, 2021). Therefore, it is crucial to balance the benefits of macro- and micro-segmentation with the need for efficient communication within the network (Hemberg et al., 2018; Katsis et al., 2021).

**Assume Breach and Plan for the Worst**

Despite the robust data security measures, a system can only be guaranteed to be utterly breach-proof if taken offline (Ghosemajumder, 2017). The reality necessitates a comprehensive approach to monitoring all access-related entities, such as data streams, devices, services, and files, collecting as much environmental information as possible to enhance the reliability of security assessment, increase the credibility organizations should conduct thorough risk assessments to identify potential threats and vulnerabilities within their infrastructure, including potential attack vectors and likelihood of successful attack (Kujo, 2023). Organizations should develop a business continuity plan outlining specific actions and protocols (e.g., establishing clear communication channels, designating responsibilities and roles, and ensuring data backup and recovery processes) to maintain critical operations during and after a breach based on the risk assessments to prepare for worst-case scenarios. The scope of risk assessment can vary depending on the specific use case, but utilizing well-prepared templates built on best practices can ensure comprehensive evaluations regardless of scope (Kujo, 2023). Additionally, a risk assessment should be carried out every time the environment changes, such as by implementing new technologies, expanding or restructuring the network, integrating third-party services, or adopting new business processes. The changes introduce new vulnerabilities and alter the risk landscape. Organizations can identify and mitigate potential risks by performing a risk assessment before they are exploited, ensuring that security measures are always aligned with the current operational environment and helping to maintain a robust security posture (National Institute of Standards and Technology, 2012).

## Issues for Implementing Zero-Trust

### Insider Threat Management

Insider threat is one of the critical risks posed by individuals within the organization (Ciampa, 2017). It is generally considered a top security concern in any organization, and also managing this inside threat is a critical component of Zero-Trust cybersecurity (Deane & Kraus, 2021). These threats can arise from current and former employees, contractors, business partners with legitimate access to the network and systems, or even cloud-computing vendors(Cappelli et al., 2012). In this situation, adopting the principle of "never trust, always verify" is essential in managing inside threats. Zero-trust cyber security can be applied effectively through continuous monitoring, strict access controls, and training for potential breaches (Ophoff et al., 2014; Rousseau, 2021).

### *Continuous Monitoring*

The first step in managing insider threats is identifying potential threats and monitoring activities within the organization. Insiders are considered "Trusted" people, so This involves deploying sophisticated tools and techniques to detect anomalies and unusual behaviors that could indicate malicious intent or risky actions (Greitzer et al., 2019). According to Greitzer et al. (2019), insider threat mitigation includes all types of technology to alert, monitor, notify, and report on activities that occur on a network. For example, this involves deploying user behavior to monitor and analyze user activities continuously. Monitoring the behaviors and ensuring that these infiltrations are unsuccessful is a behavioral analysis goal for insider threat mitigation (Homoliak et al., 2019a). It uses machine learning algorithms to establish normal behavior patterns and detect deviations that may indicate malicious intent or risky actions (Shah, 2021; Rabbani et al., 2021).

### Access Controls and Least Privilege

Implementing least-privilege security is another cornerstone of inside threat management (Ciampa, 2017; Deane & Kraus, 2021). Suppose a user has excessive privileges, or a company did not map user privileges against their actual accesses. In that case, it may be hard to identify who accessed or allowed a user to sabotage the whole system (Deane & Kraus, 2021). To reduce this problem,  well-defined Role-based access control (RBAC) ensures that users have only the permissions necessary for their roles, reducing the risk of unauthorized access (Ciampa, 2017; Deane & Kraus, 2021). Regular reviews and audits of access rights are also crucial to maintaining compliance and security (Ciampa, 2017; Deane & Kraus, 2021).

### Training and Awareness

Human error and lack of awareness are significant contributors to insider threats. Therefore, training and awareness programs are essential components of inside threat management (Ciampa, 2017; Deane & Kraus, 2021). Educating employees about the importance of cybersecurity and how to recognize and respond to potential threats can significantly reduce risks. Conducting regular training sessions on cybersecurity best practices, such as recognizing phishing attempts, proper use of passwords, and secure handling of sensitive data, can enhance employees' ability to prevent and respond to security incidents.

## Customers/Users/Patrons

Often one of the greatest potential vulnerabilities for an organization is not technology, or even internal employees, but rather the customers, users, or patrons who interact with the organization. Data is constantly being exchanged between the organization and its users. These individuals must simultaneously be viewed as subjects of cyberthreats as well as potential causes of cyberthreats – access must be balanced with ensuring control and security. Most members of the general public are woefully underprepared to prevent or address an emerging cyberthreats if targeted (Johri & Kumar, 2023; Moallem, 2019). One factor that appears important is whether the users have a stake in the technology they are using and the data that is being shared, as they are likely to be more protective of data on personal devices (Ameen et al., 2021). Another factor is whether they have ever received formal instruction about cyber security behavior, which has been shown to produce better security behavior (McCrohan et al., 2010). Transparency with users as to why new cybersecurity plans and procedures – some of which may seem inconvenient – are being implemented is key to earning buy-in (Norris et al., 2018).

### Cybersecurity Awareness for Customers/Users/Patrons

Ensuring that users are well-informed about the risks associated with interacting with the organization and systems in general is a critical dimension for shoring up this vulnerability. Training should highlight the justifications for new procedures, clearly outline what the procedures are, and provide examples and activities, as needed, to reinforce the procedures. For instance, an organization may offer exercises to highlight new procedures for multi-factor authentication or how to handle potential social engineering attacks (Miranda, 2018). Training could be formal (content the user must learn before they can access a system) or informal (reminders about best practices for cybersecurity) and may include targeted approaches aimed at particularly vulnerable user populations (Li et al., 2022). Organizations may seek to strike a balance between the potential costs of cyberthreats and the costs to train the public. The different approaches to cybersecurity awareness training that exist today provide options for the organization to consider (Zhang et al., 2021).

### *User-Focused Solutions*

Users are not experts in cybersecurity. They very well may not even be familiar with the systems in use within an organization. Thus, it is critical to meet the users where they are at in terms of knowledge and ability. Systems that are designed with high usability are less likely to engage in system behavior that could leave the organization susceptible to threats (Nurse et al., 2011). Clear policy surrounding the use of these systems will further support these security initiatives (AlQadheeb et al., 2022). A Zero-Trust architecture will naturally support greater security, but it too needs buy-in, given the amount of change it may require (Phiayura & Teerakanok, 2023). Incorporating feedback from actual users may help support this process.

### Hybrid Cloud Protection

Hybrid cloud environments, which integrate on-premises infrastructure with public and private cloud services, pose unique security challenges. Implementing Zero-Trust in these settings necessitates a comprehensive strategy to protect data, applications, and infrastructure across all platforms. The Zero-Trust Cybersecurity principle of "never trust, always verify" is essential for securing hybrid clouds. Unlike on-premises infrastructure, cloud systems are external to the organization. Numerous stakeholders and components are involved between the cloud vendor and the organization utilizing the cloud service.

### *Challenges in Hybrid Cloud Security*

While hybrid cloud environments offer flexibility and scalability, they also introduce complexities in security management. Many researchers have identified the biggest challenges in cloud computing, including secure data storage, data confidentiality, accountability, and data encryption), high-speed access to the Internet, lack of audit features, portability, interoperability, linkage, organizational sustainability, and standardization, as well as maintaining consistent security policies across different environments, managing access controls, securing data both in transit and at rest, and ensuring visibility, privacy, and control over all components (Tissir et al., 2021). Tabrizchi and Rafsanjani (2020) emphasizes that the major challenge in the adoption of the cloud is security and these challenges require a robust security framework that integrates seamlessly across on-premises and cloud platforms, ensuring consistent policy enforcement and threat detection. The "never trust, always verify" principle is especially vital in hybrid cloud environments. Every access request, regardless of its origin, must be continuously authenticated and authorized. The dynamic nature of hybrid clouds, characterized by frequent data transfers between on-premises and cloud services, complicates security management further. To address these challenges effectively, organizations must adopt advanced technologies and strategies (Cloud Security Alliance, 2021).

### Implementing a Security Framework

Adopting a robust security framework is essential for deploying Zero-Trust in hybrid cloud environments. This framework should incorporate security policies, processes, and technologies to ensure uniform protection across all platforms. The National Institute of Standards and Technology (NIST) offers a comprehensive framework for managing cybersecurity risks, suitable for hybrid cloud settings. The NIST Cybersecurity Framework includes six core functions: Govern, Identify, Protect, Detect, Respond, and Recover. These functions offer a structured approach to managing cybersecurity risks and applying Zero-Trust principles (NIST, 2024). Organizations should adapt the NIST framework to their unique needs, integrating best practices and industry standards (Ciampa, 2017; Deane & Kraus, 2021). In addition, Service level agreements (SLAs) play a crucial role in managing security in hybrid cloud environments by defining the performance and security standards that cloud service providers must meet. These agreements help manage customer expectations and outline the circumstances under which providers are not liable for outages or performance issues. While SLAs represent the performance characteristics of services and facilitate comparisons, they do not guarantee service quality or eliminate the risk of selecting a poor service provider. Typically, an SLA includes a statement of objectives, a list of services covered, and the responsibilities of both the service provider and the customer. Key considerations in SLAs include robust data encryption, comprehensive monitoring and threat detection capabilities, clearly defined incident response protocols, and compliance with industry standards and regulatory requirements. By including these provisions, organizations can ensure that their hybrid cloud environments remain secure and that service providers are held accountable for their security practices (Fotiou et al., 2015).

### Data Protection

Cloud computing environments are characterized by diverse, sparsely distributed nodes that are often difficult to control effectively. Data protection in cloud computing encompasses several critical areas, including encryption, access control, and trust management (Sun, 2020). Data encryption is vital for ensuring the confidentiality and integrity of data in hybrid cloud environments. Encrypting data in transit and at rest keeps sensitive information secure even if intercepted or accessed without authorization (Ciampa, 2017). While cloud providers offer various encryption services, organizations must securely manage encryption keys to prevent unauthorized access. Implementing end-to-end encryption, where data is encrypted before transmission and decrypted only by authorized recipients, further enhances security (Deane & Kraus, 2021). The integration of encryption and access control with trust management further enhances data protection in cloud computing. Trust models can be incorporated into encryption schemes to evaluate and ensure the reliability of users and service providers. This holistic approach to data protection not only secures data but also maintains the integrity and availability of cloud services, thereby fostering a trustworthy computing environment (Sun, 2020).

### Privacy

Cloud computing environments face numerous privacy challenges due to the technology's inherent nature. The risk of privacy breaches escalates as data is transmitted, processed, and stored by cloud service providers (CSPs). This occurs because any security vulnerabilities in existing technologies are carried over to the cloud platform, amplifying potential security threats (Sun, 2020). According to the Cloud Security Alliance (CSA) and other scholarly sources, key privacy threats in cloud computing include data disclosure, access rights management issues, and difficulties in data destruction (Reza & Satyajayant, 2018). Virtualization in cloud services also introduces privacy challenges, as attacks among virtual machines can happen despite isolation strategies. Additionally, multi-tenant and cross-domain sharing can complicate service authorization and access control, increasing the risk of unauthorized access and data breaches (Sun, 2020; Yang, 2021). Addressing privacy concerns in cloud computing requires a multi-faceted approach that incorporates advanced technologies, robust policies, and best practices (Kumar et al., 2018; Reed et al., 2011). According to Kumar et al., (2018) that reducing data privacy problems, organizations should ensure they know the logical and physical location of their data, including the state, country, and specific data center, to address potential regulatory, contractual, and jurisdictional issues; Establishing location and jurisdictional policies to govern data location is essential; Intelligent data segregation techniques should be adopted to separate data from different users effectively; using strong encryption techniques for backup data is crucial to prevent data leakage.

### Monitoring and Detection

Comprehensive visibility and real-time threat detection are crucial for managing security in any networked system including hybrid cloud environments (NIST, 2024; Ciampa, 2017; Deane & Kraus, 2021). Continuous monitoring and advanced detection capabilities allow organizations to promptly identify and respond to threats. For example, Security Information and Event Management (SIEM) systems are vital in this process (Deane & Kraus, 2021). SIEM solutions collect and analyze logs and events from various sources within the hybrid cloud environment, providing a complete view of security activities. By correlating events and identifying patterns indicative of malicious behavior, SIEM systems facilitate rapid threat detection and response. Cloud Security Posture Management (CSPM) tools automate the assessment of cloud security configurations, identifying potential vulnerabilities and ensuring compliance with security policies (Loaiza Enriquez, 2021). CSPM solutions continuously monitor cloud environments for misconfigurations and deviations from best practices, helping organizations maintain a secure posture.

### Access Controls and Least Privilege

Implementing access controls and the principle of least privilege is fundamental to securing cloud environments. This approach involves granting users only the permissions necessary to perform their tasks, thus minimizing the risk of unauthorized access and data breaches. To ensure data confidentiality and integrity, organizations must utilize advanced encryption techniques for data in transit and at rest, while securely managing encryption keys through key management services (Deane & Kraus, 2021; Sun, 2020). In addition, advanced authentication protocols such as anonymous two-factor user authentication and dynamic reciprocal authentication offer secure mutual authentication, protecting against phishing and man-in-the-middle attacks (Mo et al., 2020; Ahmed et al., 2021). Smart virtual cards and blockchain technology provide additional layers of security by ensuring the integrity and authenticity of transactions (Derhab et al., 2020). Furthermore, machine learning-based intrusion detection systems, like those using support vector machines (SVM) and information gain (IG), improve the accuracy and speed of detecting malicious activities, thereby enhancing overall cloud security (Mugabo et al., 2020). The Mobile Cloud Intrusion Detection and Prevention System (MINDPRES) leverages machine learning to dynamically analyze network traffic and device resources, providing robust protection against intrusions (Ogwara et al., 2021). These combined strategies form a comprehensive approach to securing cloud environments and protecting sensitive data.

## Contextual Differences in Zero-Trust Cybersecurity

Importantly, the appearance of zero-trust implementation may differ based on context. A for-profit organization with few customers will look distinct from a public library. Both organizations need a high-level of security as common targets of attacks, but the threats for a small organization with few customers (likely external threat) are different from those of a library with many public users. Additionally, the targets of attacks may differ. A for-profit organization may be attacked for financial information, whereas a public library may be attacked for patron data or to hijack systems for ransom. These factors are all important in the design of the zero-trust architecture. The following sections explore several unique contexts in detail.

### The University Environment

Institutions of Higher Education (IHEs) hold access to protected information not only about employees (e.g., social security numbers), but also thousands of students. A breach of this information could not only cause irreparable damage to the reputation of the institution and put students and employees at risk but make the institution criminally liable for failing to protect these parties' information (Jackson, 2021). Obviously, this would come with severe direct and indirect impacts on the institution's financial standing and public trust. Given these consequences, preserving security at all costs is vital.

Many colleges and universities already utilize strategies like multi-factor authentication to prevent hacking, but one can argue that these measures are insufficient. There are many systems within universities that hold very sensitive information and yet are accessible to lower-level employees like part-time and student workers (Ghosh et al., 2016). Employees at all levels regularly access systems from different locations around campus – an instructor could easily forget to log out of a classroom computer station. The vulnerabilities are practically boundless. Zero-trust solutions may provide an answer to protect these valuable higher education resources.

Here are a few examples of how zero-trust can support cybersecurity in higher education:

- Zero-trust can limit access to only the information employees need, when they need it (DeWeaver, 2021). For instance, it is possible a student employee may need to access student records in the course of their work, but they have no legitimate rationale to have access to this information outside of work hours and their workstation.

- Faculty members have substantial amounts of information, including student grades and funding members, that must be protected (Culnan & Carlin, 2009). When they leave a computer station unattended – such as in a classroom when they leave to use the restroom – they create a vulnerability. Session timeouts can protect these workstations by locking the computer and requiring a fresh log-in to access the station again. While this solution may cause frustration for some faculty members, it may also prevent a major breach.
- Students require access to many systems, offering a slightly different dynamic where they must share large amounts of private information but have limited access to the stored information of others (Daraghmi et al., 2019). Permissions must be managed to protect students from their aown peers.

**The Library Environment**

Libraries hold immense stores of information in the form of the copyrighted physical and digital works they lend to patrons, the access they afford to the Internet, and the data they possess about their patrons (Lund, 2021). All of this information is potentially valuable to attackers. If, for instance, a hacker gains access to patrons' sensitive information, they could hold it for ransom, like with Toronto Public Libraries. As with institutions of higher education, libraries present unique challenges by having both employee and patron/user populations to manage as far as cyberthreats (Hess et al., 2015).

Within libraries, patrons must have access to their own data and data about library resources, but not data pertaining to other patrons. Front-line library workers must have some ability to look up information but do not usually need access to information about other aspects of internal library operations or fellow employees. Administrators, however, need access to wide ranging data. This necessitates varying levels of permissions based on an individual's credentials (Amini et al., 2021). Fortunately, this mandate is built directly into zero-trust cybersecurity. Additional ways that zero-trust may support cybersecurity in libraries include:

- Protecting patrons against invasions of privacy by authorities could be supported by zero-trust measures. Historically, library records have been a target of police, who might use them to monitor patron behavior. The American Library Association, the leading organization for libraries, strongly opposes this activity and supports practices that restrict these efforts (Mars, 2017). Nonetheless, it can be intimidating for an unprepared front-line library worker if confronted by law enforcement. A zero-trust system could prevent these officials from easily gaining access to this information from a front-line employee, forcing them to follow the prescribed path of receiving a warrant and communicating with the library director.
- As with the case of an instructor who leaves a computer unattended, session timeouts can be used to secure employee workstations to ensure no unmonitored patrons gain access to unauthorized information (Dietz, 2022).

**The Supply Chain Environment**

A supply chain encompasses entities directly providing and distributing products, services, funds, and information from origin to destination (Mentzer et al., 2001). In contemporary society, supply chains are integral to daily life, facilitating the delivery of essential items such as water, food, healthcare, medications, and energy resources (Council of Supply Chain Management Professionals, n.d.). Supply Chain Management covers critical functions, including comprehensive planning, sourcing, production, delivery, and returns management (Felea & Albăstroiu, 2013). However, the extensive interconnections among stakeholders, technologies, and geographic locations in contemporary supply chain systems introduce vulnerabilities malicious actors can exploit (Canadian Centre for Cyber Security, 2022). Additionally, the emergence of big data has led to exponential growth in data across the supply chain, encompassing information from procurement, production, distribution, and customer interaction, which increases data security complexity (Gopal et al., 2024). Big data analytics provides valuable insights for optimizing operations, predicting demand, and enhancing the customer experience. Meanwhile, it broadens the attack surface and makes organizations vulnerable to potential breaches, particularly identity-based attacks such as theft or misuse of user credentials, privileges, or personal information. The vast amount of data flowing through the supply chain can be challenging to monitor and secure, and participants accessing large data sets may inadvertently or intentionally misuse or disclose sensitive information, exacerbating security challenges (Ogbuke et al., 2022).

Achieving zero-trust in the supply chain involves developing comprehensive, enterprise-wide security plans and strategies (Collier & Sarkis, 2021). The plans address the intricate relationships between upstream and downstream stakeholders, flows of material, information, and finances, and access transaction strategies. Unlike the IT field, the supply chain encompasses technical systems and complex processes, individuals, and relationships, all requiring careful consideration and attention (Collier & Sarkis, 2021). Based on the National Institute of Standards and Technology (2020) zero-trsut architecture guideline, Collier and Sarkis (2011) proposed the following transitional steps for implementing zero-trust in the supply chain:

- Supply chain organizations need to identify participants and boundaries, distinguish between internal and external participants (including suppliers, clients, and internal employees), and understand their roles and the level of access required (Collier & Sarkis, 2021).
- Identify supply chain assets by cataloging data, information, and systems within the enterprise, recognizing non-enterprise participants and technologies that interact with the supply chain, and understanding general business processes related to the organization's mission, such as trust-related processes and contractually mandated procedures for non-enterprise participants, identifying threats posed by participants, assets, and processes, and conducting risk assessments to prioritize zero-trust implementation and its impact on business objectives ((Collier & Sarkis, 2021).
- During deployment and monitoring, the organization should decide on a deployment strategy, possibly using a trial mode, and gather the necessary data to evaluate success while ensuring the ability to revert to the previous configuration (Collier & Sarkis, 2021).
- Finally, implementing zero-trust involves designing an iterative process that builds on successes and learns from failures, gradually transitioning, adjusting priorities, and incorporating continuous improvement into deployments (Collier & Sarkis, 2021).

## Conclusion

Zero-trust cybersecurity has proven to be an essential framework for addressing the complex and evolving threat landscape organizations face today. By shifting from a traditional "trust but verify" approach to "never trust, always verify," zero-trust emphasizes continuous verification, strict access controls, and the assumption that breaches are inevitable (Buck et al., 2021). This paper has explored the core principles of zero-trust, including continuous authentication, least privilege security, and breach assumption. Specific steps for implementing zero-trust, with a detailed focus on inside threat management and hybrid cloud protection, have been examined, alongside the unique challenges and strategies for different environments such as libraries, universities, and warehouses.

Understanding contextual differences in Zero-Trust cybersecurity is essential for effective implementation. In university environments, where diverse user groups and open networks are prevalent, securing research data, protecting student information, and ensuring secure access to educational resources is paramount. In the library environment, the focus should be on protecting patron data, securing public access computers, and ensuring the integrity of digital resources. Warehouses, relying heavily on automation and IoT devices, require strategies to secure IoT devices, protect inventory data, and ensure physical security. By tailoring Zero-Trust strategies to these specific environments, organizations can enhance their security posture and better protect their data, systems, and users from evolving cyber threats. While integrating these principles may be frustrating at first for some workers accustomed to fewer restrictions, protecting one's organization is of paramount importance.

Future research may focus on developing advanced, user-friendly security analytics tools that leverage artificial intelligence and machine learning for real-time threat detection and response. More efficient and cost-effective Zero-Trust implementation methods should be explored, especially for small and medium-sized enterprises. Specific Zero-Trust strategies tailored to data-rich and information-rich contexts is necessary. The evolving landscape of remote work and hybrid cloud environments also warrants ongoing research to identify best practices and innovative solutions. Another critical area for future study is to balance security measures with usability to ensure that security protocols do not hinder organizational efficiency or user satisfaction. Longitudinal studies examining Zero-Trust implementations' long-term effectiveness and adaptability across various industries would provide valuable insights into the model's sustainability.